\documentclass[12pt]{article}
\usepackage{putex}
\usepackage{graphicx}
\begin{document}

\preprint{hep-th/0605292 \\ PUPT-2199}

\institution{PU}{Joseph Henry Laboratories, Princeton University, Princeton, NJ 08544}

\title{Dissipation from a heavy quark moving through ${\cal N}=4$ super-Yang-Mills plasma}

\authors{Joshua J. Friess, Steven S. Gubser, and Georgios Michalogiorgakis}

\abstract{Using AdS/CFT, we compute the Fourier space profile of
$\langle \tr F^2 \rangle$ generated by a heavy quark moving through
a thermal plasma of strongly coupled ${\cal N}=4$ super-Yang-Mills
theory.  We find evidence of directional emission from the quark whose description includes
gauge fields with large momenta.  We comment on the possible
relevance of our results to relativistic heavy ion collisions.}


\date{May 2006}

\maketitle

\section{Introduction}
\label{INTRODUCTION}

In \cite{Herzog:2006gh,Gubser:2006bz}, a classical solution of string theory is described that is dual in the sense of AdS/CFT \cite{Maldacena:1997re,Gubser:1998bc,Witten:1998qj} to an external quark passing through a thermal plasma of ${\cal N}=4$ super-Yang-Mills theory at large $N$ and strong 't~Hooft coupling $g_{YM}^2 N$.  The string is treated in the test string approximation: its back-reaction on the geometry is not considered.  The string dangles into $AdS_5$-Schwarzschild from an external quark on the boundary which is constrained to move with constant velocity.  The string trails out behind the quark and exerts a drag force
 \eqn{DragForce}{
  {dp \over dt} =
    -{\pi \sqrt{g_{YM}^2 N} \over 2} T^2 {v \over \sqrt{1-v^2}}
 }
on the quark.  Here $v$ is the speed of the quark, and $T$ is the temperature of the plasma, or equivalently the Hawking temperature of the horizon of $AdS_5$-Schwarzschild.  The diffusion constant $D = 2/(\pi T \sqrt{g_{YM}^2 N})$ implied by \eno{DragForce} was derived independently in \cite{Casalderrey-Solana:2006rq}, also using AdS/CFT.

In the gauge theory, energy loss results from gluons (or superpartners of gluons) radiating off the heavy quark and interacting with the plasma.  We should ask: How energetic are these radiated gluons?  At what angle do they come off relative to the velocity of the heavy quark?  To the extent that such questions can be posed in a gauge-invariant manner, AdS/CFT should be able to provide an answer.  The aim of the present paper is to shed some light on these questions by computing the profile of $\langle \tr F^2 \rangle$ in the boundary gauge theory.  To do this we compute the linear response of the dilaton field to the string, which is a first step in computing its back-reaction on the $AdS_5$-Schwarzschild background.  Actually, what we will extract in the end is the vacuum expectation value (VEV) of the operator in ${\cal N}=4$ super-Yang-Mills which couples to the dilaton.  This is not quite $\tr F^2$, but rather the lagrangian density plus a total derivative: in mostly plus signature,
 \eqn{CalO}{
  {\cal O}_{F^2} = {1 \over 2 g_{YM}^2} \tr\left( -F_{mn}^2 +
    2 X^I D_m^2 X^I - 2i \bar\lambda^a \bar\sigma^m D_m \lambda_a +
    {\rm more\ interactions} \right) \,,
 }
where $X^I$ are the six adjoint scalars, $\lambda_a$ are the four Weyl adjoint fermions, $D_m$ is the gauge-covariant derivative, and $\bar\sigma^m = (-{\bf 1},-\vec\sigma)$ where $\vec\sigma$ are the Pauli matrices.

The near field of the heavy quark is just the Coulomb color-electric flux, appropriately Lorentz boosted.  The contribution of this near field to $\langle {\cal O}_{F^2} \rangle$ can be computed analytically, following \cite{Danielsson:1998wt}, and it has nothing to do with energy loss.  When it is subtracted away from $\langle {\cal O}_{F^2} \rangle$, the remainder is peaked at momenta many times
larger than the temperature.  The
information in $\langle {\cal O}_{F^2} \rangle$ is complementary to
\eno{DragForce} in that it helps identify the energy scale at which
dissipative phenomena occur but does not so clearly indicate the
overall rate of dissipation. More complete information could be
extracted from $\langle T_{\mu\nu} \rangle$, which could also be
computed via AdS/CFT but requires a more technically involved
treatment of metric perturbations.

Several related papers \cite{Liu:2006ug,Buchel:2006bv,Herzog:2006se,Caceres:2006dj} appeared recently, all aiming to describe at some level energy dissipation from a fundamental quark into a thermal plasma using AdS/CFT.  The interest in this topic owes to a possible connection with relativistic heavy ion physics.  A distinctive feature of RHIC experiments \cite{Arsene:2004fa,Adcox:2004mh,Back:2004je,Adams:2005dq} is jet-quenching, which is understood as strong energy loss as a high-energy parton passes through the quark-gluon plasma formed in a gold-on-gold collision.

The organization of the rest of this paper is as follows.  In
section~\ref{DILATON} we explain the classical supergravity
calculation that leads to $\langle {\cal O}_{F^2} \rangle$.  Similar calculations were carried out in \cite{Callan:1999ki} for a string undergoing small oscillations around certain static configurations in $AdS_5$.  All the supergravity
computations are done in five dimensions, but the final answer is the gauge theory quantity
$\langle {\cal O}_{F^2} \rangle$ as a function of the coordinates
$(t,x^1,x^2,x^3)$ of Minkowski space.  (Actually we will find it
easier to pass to momentum space early in the computation.)  One must solve a boundary value problem in order to extract $\langle {\cal O}_{F^2}
\rangle$.  Numerical techniques for doing so and results for several
different choices of $v$ are described in section~\ref{NUMERICS}.
We conclude in section~\ref{DISCUSSION} with a discussion of the
possible relevance of our work to recent experimental results.

\section{Dilaton perturbations}
\label{DILATON}

The background geometry is the well-known $AdS_5$-Schwarzschild solution,
 \eqn{AdSSch}{
  ds^2 = G_{\mu\nu} dx^\mu dx^\nu =
    {L^2 \over z^2} (-h dt^2 + d\vec{x}^2 + dz^2/h) \qquad
      h = 1 - {z^4 \over z_H^4} \,,
 }
and useful relations include
 \eqn{TtHooft}{
  {L^4 \over \alpha'^2} = g_{YM}^2 N \qquad
    T = {1 \over \pi z_H} \,.
 }
In static gauge, the string worldsheet is described as
 \eqn[c]{Xstatic}{
  X^\mu(t,z) \equiv
   \begin{pmatrix} t & X^1(t,z) & 0 & 0 &
     z \end{pmatrix}  \cr
  X^1(t,z) = vt + \xi(z) \qquad
  \xi(z) = -{z_H v \over 4i} \left( \log {1-iz/z_H \over 1+iz/z_H} +
    i \log {1+z/z_H \over 1-z/z_H} \right) \,.
 }

To compute the dilaton response to this string, one starts with the following action:
 \eqn{SUGRAandString}{
  S = \int d^5 x \, \sqrt{-G} \left[ -{1 \over 4\kappa_5^2}
    (\partial\phi)^2 \right] -
   {1 \over 2\pi\alpha'} \int_M d^2\sigma \, e^{\phi/2} \sqrt{-g}
   \qquad g_{\alpha\beta} \equiv G_{\mu\nu} \partial_\alpha X^\mu
    \partial_\beta X^\nu \,.
 }
Here $\alpha$ and $\beta$ refer to worldsheet coordinates $\sigma^\alpha = (\tau,\sigma)$, and
 \eqn{VarKappaDef}{
  \kappa_5^2 = {4\pi^2 L^3 \over N^2} = 8\pi G_5
 }
where $G_5$ is the five-dimensional gravitational constant.  To derive the dilaton equation of motion, it helps first to rewrite the whole action as a single volume integral (we refrain briefly from choosing static gauge):
 \eqn{SingleIntegral}{
  S = \int d^5 x \, \sqrt{-G} \left[
    -{1 \over 4\kappa_5^2} (\partial\phi)^2 -
    {1 \over 2\pi\alpha'} \int d^2 \sigma \, e^{\phi/2}
     {\sqrt{-g} \over \sqrt{-G}} \delta^5(x^\mu-X^\mu(\sigma))
    \right] \,.
 }
The five-dimensional delta function in \eno{SingleIntegral} is a product of standard Dirac delta functions.  So, for instance,
 \eqn{ProductDelta}{
  \delta^5(x^\mu) = \delta(t) \delta(x^1) \delta(x^2) \delta(x^3)
    \delta(z) \,.
 }
The linearized equation of motion can now be straightforwardly derived as
 \eqn{PhiEOM}{
  \square\phi = {1 \over \sqrt{-G}} \partial_\mu \sqrt{-G}
    G^{\mu\nu} \partial_\nu \phi = J \equiv
   {\kappa_5^2 \over 2\pi\alpha'} \int d^2 \sigma \,
    {\sqrt{-g} \over \sqrt{-G}} \delta^5(x^\mu-X^\mu(\sigma)) \,.
 }
and by passing to static gauge one may explicitly perform then the remaining integral in \eno{PhiEOM}:
 \eqn{StaticJ}{
  J &= {\kappa_5^2 \over 2\pi\alpha'} {\sqrt{-g} \over \sqrt{-G}}
     \delta(x^1-X^1(t,z)) \delta(x^2) \delta(x^3) \,.
 }

In the spirit of finding the steady-state, late-time behavior, we assume that $\phi$ depends on $x^1$ and $t$ only through the combination $x^1-vt$.  After computing
 \eqn{SqrtComputation}{
  {\sqrt{-g} \over \sqrt{-G}} = {z^3 \over L^3} \sqrt{1-v^2} \,,
 }
one can easily show that $\square \phi = J$ simplifies to
 \eqn{BestPDE}{
  \left[ z^3 \partial_z {h \over z^3} \partial_z +
    \left( 1 - {v^2 \over h} \right) \partial_1^2 +
     \partial_2^2 + \partial_3^2 \right] \phi =
    {\kappa_5^2 \sqrt{1-v^2} \over 2\pi\alpha'} {z \over L}
     \delta(x^1 - vt - \xi(z)) \delta(x^2) \delta(x^3) \,.
 }
This partial differential equation can be attacked by Fourier transforming:
 \eqn{ThreeFourier}{
  \phi(t,\vec{x},z) =
    \int {d^3 k \over (2\pi)^3} \,
     e^{i k_1 (x^1-vt) + ik_2 x^2 + ik_3 x^3}
      \phi_k(z) \,,
 }
and similarly for $J$.  Then one has
 \eqn{BestODE}{
  \left[ z^3 \partial_z {h \over z^3} \partial_z -
    \left( 1 - {v^2 \over h} \right) k_1^2 - k_\perp^2 \right]
   \phi_k
   = {\kappa_5^2 \sqrt{1-v^2} \over 2\pi\alpha'} {z \over L}
      e^{-i k_1 \xi(z)} \,,
 }
where $k_\perp^2 = k_2^2 + k_3^2$.  All dimensionful factors drop out of the differential equation when we introduce rescaled variables
 \eqn{RescaledVars}{
  K_1 = z_H k_1 \qquad
   K_\perp = z_H k_\perp \qquad
   y = {z \over z_H} \qquad
   \tilde\phi_K(y) = {2\pi\alpha' L \over \kappa_5^2 z_H^3}
     {1 \over \sqrt{1-v^2}} \phi_k(z) \,.
 }
Then $h=1-y^4$ and
 \eqn{RescaledEOM}{
  \left[ y^3 \partial_y {h \over y^3} \partial_y -
    \left( 1 - {v^2 \over h} \right) K_1^2 - K_\perp^2 \right]
     \tilde\phi_K = y e^{-iK_1 \xi/z_H} =
     y \left( {1-iy \over 1+iy} \right)^{vK_1/4}
      \left( {1+y \over 1-y} \right)^{ivK_1/4} \,.
 }
There doesn't appear to be a solution to \eno{RescaledEOM} in terms of known special functions.  However it can be solved in two interesting limiting regimes:
 \begin{itemize}
  \item {\bf Near the horizon,} $y$ is slightly less than $1$, a better choice of radial variable is $Y = \log(1-y)$.  The leading terms in the differential equation near the horizon (that is, for large negative $Y$) are
 \eqn{LeadingHorizon}{
  \left[ \partial_Y^2 + \left( vK_1 \over 4 \right)^2 \right]
   \tilde\phi_K = {1 \over 4} e^Y e^{-ivK_1 (Y+\pi/2 - \log 2)/4} \,,
 }
which is also the equation of motion for a simple harmonic oscillator with a complex driving force.  The solutions are
 \eqn{HorizonSolns}{
  \tilde\phi_{{\rm near},K} = {e^Y/4 \over 1-ivK_1/2}
    e^{-ivK_1(Y+\pi/2-\log 2)/4} +
   C_K^+ e^{ivK_1Y/4} +
   C_K^- e^{-ivK_1Y/4} \,,
 }
where $C_K^\pm$ are arbitrary constants.  The standard boundary condition at a black hole horizon is to choose a purely infalling solution.  This means that in the near-horizon limit, $\phi$ should depend on $t$ and $Y$ only through the combination $t+z_H Y/4$, not $t-z_H Y/4$: the quantity $z_H Y/4$ is essentially the tortoise coordinate.  Thus $C_K^+ = 0$.
  \item {\bf Near the boundary} of $AdS_5$-Schwarzschild, the leading terms in the differential equation are
 \eqn{LeadingBdy}{
  y^3 \partial_y {1 \over y^3} \partial_y \tilde\phi_K = y \,,
 }
and the solutions are
 \eqn{BdySolns}{
  \tilde\phi_{{\rm far},K} = -{y^3 \over 3} + A_K + B_K y^4 \,,
 }
where $A_K$ and $B_K$ are arbitrary constants.  $A_K$ should be set to zero because there is no deformation of the lagrangian.  $B_K$ is proportional to $\langle {\cal O}_{F^2} \rangle$.
 \end{itemize}

It is worth noting that the relation of $B_K$ to $\langle {\cal O}_{F^2} \rangle$ involves a subtraction of contact terms.  Conventionally, it is understood that
 \eqn{GotFSquared}{
  \langle {\cal O}_{F^2}(t,\vec{x}) \rangle = -{L^3 \over 2\kappa_5^2}
    \lim_{z\to 0} {1 \over z^3} \partial_z \phi(t,\vec{x},z) \,,
 }
but in the present case, the limit doesn't exist because of the $y^3$ term in \eno{BdySolns}.  Fortunately, this term has no $\vec{K}$ dependence.  Thus when passing back to real space, it is proportional to a delta function supported at the location of the quark.  This delta function has an infinite coefficient, but if it is subtracted, the remaining contribution to $\langle {\cal O}_{F^2}(t,\vec{x}) \rangle$ indeed comes from $B_K$, and it is finite.  The subtraction prescription has some arbitrariness: one could subtract off any finite multiple of the delta function at the same time, which corresponds to subtracting a $K$-independent quantity from every $B_K$.

Combining \eno{TtHooft}, \eno{ThreeFourier}, \eno{RescaledVars}, \eno{BdySolns}, and \eno{GotFSquared}, one finds
 \eqn{GotO}{
  \langle {\cal O}_{F^2}(t,\vec{x}) \rangle =
    -\pi^3 T^4 \sqrt{g_{YM}^2 N} \sqrt{1-v^2}
     \int {d^3 K \over (2\pi)^3} e^{\left[ iK_1(x^1-vt) +
       i K_2 x^2 + i K_3 x^3 \right] / z_H} B_K \,.
 }
In section~\ref{NUMERICS}, we will quote results in units where $z_H = 1$: this corresponds to $T = 1/\pi$.

For a wide range of $K_1$ and $K_\perp$, the dominant contribution
to $B_K$ comes from the near field of the quark, which in position
space is proportional to $1/|\vec{x}|^4$ in the rest frame of the
quark.  Consider first the case $v=0$.  Following \cite{Danielsson:1998wt}, consider a string dangling straight down in $AdS_5$.  One obtains
 \eqn{NearFieldDan}{
  \langle {\cal O}_{F^2}(t,\vec{x}) \rangle =
    {1 \over 16 \pi^2} {\sqrt{g_{YM}^2 N} \over |\vec{x}|^4} \,.
 }
This calculation is done in the absence of a horizon, or equivalently at zero temperature.  Fourier transforming \eno{NearFieldDan} leads to
 \eqn{FoundBK}{
  B^{\rm near\ field}_K = {\pi \over 16} |\vec{K}| =
    {\pi \over 16} \sqrt{K_1^2 + K_\perp^2} \,.
 }
We have expressed the result in terms of the dimensionless variables \eno{RescaledVars} with $z_H=1/\pi T$ finite, even though $T=0$ physically.  This is a bookkeeping trick to obtain a form that can easily be compared with $AdS_5$-Schwarzschild results.

For $v \neq 0$, one may apply a Lorentz boost to the $AdS_5$ string configuration considered in the previous paragraph.  This describes an external quark moving through the vacuum at speed $v$.  The result for $B^{\rm near\ field}_K$ in this case is
 \eqn{FoundBKagain}{
  B^{\rm near\ field}_K =
    {\pi \over 16} \sqrt{(1-v^2) K_1^2 + K_\perp^2} \,.
 }
This is the analytic form that we will subtract from numerically evaluated $B_K$ to excise the near field but leave behind all the dissipative dynamics.

\section{Numerical algorithms and results}
\label{NUMERICS}

The boundary value problem described in and below \eno{RescaledEOM}  is reminiscent of both the glueball calculations initiated in \cite{Witten:1998zw,Csaki:1998qr} and of quasi-normal modes in $AdS_5$-Schwarzschild \cite{Horowitz:1999jd}.  But there is an additional simplifying feature: all the equations are affine in $\tilde\phi_K$---that is, they are linear combinations of $\tilde\phi_K(y)$, its derivatives, and functions of $y$ that do not involve $\tilde\phi_K(y)$.  To see this, consider the following formulation of the horizon boundary condition.  One first expresses the asymptotic solutions $\tilde\phi_{{\rm near},K}$ and $\tilde\phi_{{\rm far},K}$ as a sum of the inhomogenous solution and the permitted homogenous solution.  Explicitly, for the near-horizon solution,
 \eqn{NearSeparate}{
  \tilde\phi_{{\rm near},K} &= \tilde\phi_{{\rm near,P},K} +
    C_K^- \tilde\phi_{{\rm near,H},K}  \cr
  \tilde\phi_{{\rm near,P},K} &\equiv {e^Y/4 \over 1-ivK_1/2}
    e^{-ivK_1(Y+\pi/2-\log 2)/4}  \cr
  \tilde\phi_{{\rm near,H},K} &\equiv e^{-ivK_1Y/4} \,.
 }
The Wronskian
 \eqn{NearWron}{
  W_{\rm near}(y) =
   (\tilde\phi_K(y) - \tilde\phi_{{\rm near,P},K}(y))
    \tilde\phi'_{{\rm near,H},K}(y) -
   (\tilde\phi_K'(y) - \tilde\phi'_{{\rm near,P},K}(y))
    \tilde\phi_{{\rm near,H},K}(y)
 }
is a measure of how close the numerically computed function $\tilde\phi_K(y)$ is to the analytic approximation $\tilde\phi_{{\rm near},K}$.  Because the horizon is a singular point of the differential equation, one must impose the boundary condition $W_{\rm near}(y_1)=0$ at a point $y_1$ slightly less than $1$, which is to say slightly outside the horizon.  The quantity $W_{\rm near}(y_1)$ is indeed a linear combination of $\tilde\phi_K(y)$, $\tilde\phi_K'(y)$, and a $\tilde\phi_K$-independent function known in terms of $\tilde\phi_{{\rm near,H},K}(y)$ and $\tilde\phi_{{\rm near,P},K}(y)$.  One may similarly formulate a boundary condition $W_{\rm far}(y_0)=0$ which is also affine in $\tilde\phi_K$.  The point $y_0$ should be chosen slightly greater than $0$, which is to say close to the boundary of $AdS_5$-Schwarzschild.

\begin{figure}
  \centerline{\includegraphics[height=7in]{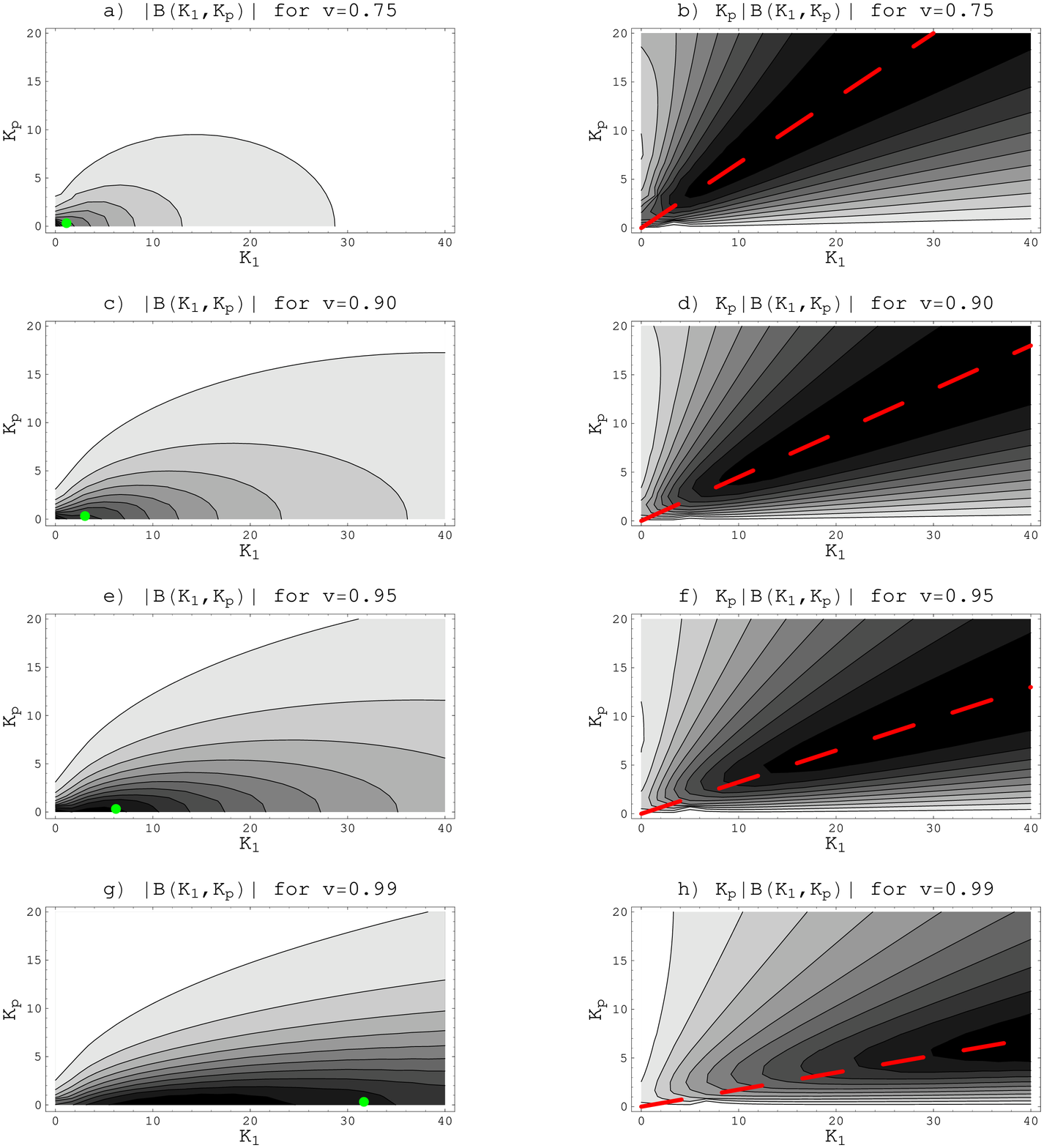}}
  \caption{The absolute value of $B(K_1,K_\perp)$ with and without the phase space factor $K_\perp$.  The near field contribution \eno{FoundBKagain} has been subtracted.  The green dot is the recoil energy of a thermal gluon: see \eno{TGRE}.  The dashed red lines indicate the direction in which $K_\perp |B(K_1,K_\perp)|$ is largest: see the discussion around \eno{ThetaTable}.  In each plot, the white region is closest to zero, and the black region is the most positive.}\label{AbsValues}
 \end{figure}

There are special methods to solve boundary value problems of the
type just described, where both the differential equation and the
boundary conditions are affine, which are more efficient than
standard shooting algorithms.  Mathematica's {\tt NDSolve}
incorporates such methods internally \cite{ChasingMethods}.  But we
have found that we achieve greater numerical accuracy using a
home-grown shooting method where $B_K$ is guessed and then adjusted
to make $C_K^+=0$.  Accuracy was further improved by finding power
series corrections to the asymptotic forms \eno{HorizonSolns} and \eno{BdySolns}.  A
satisfactory choice of cutoff points was $y_0=0.01$ and $y_1 =
0.99$.  The numerical challenge increases as $K_1$ and $K_\perp$
increase, requiring more CPU time.  As we will see in figure~\ref{AbsValues}, $B_K$ is significantly weighted toward $K$ larger than $10$ when an appropriate phase space factor is included.  So it would be
worthwhile to have some alternative method adapted to this regime,
perhaps based on a WKB approximation.

\begin{figure}
  \centerline{\includegraphics[height=7in]{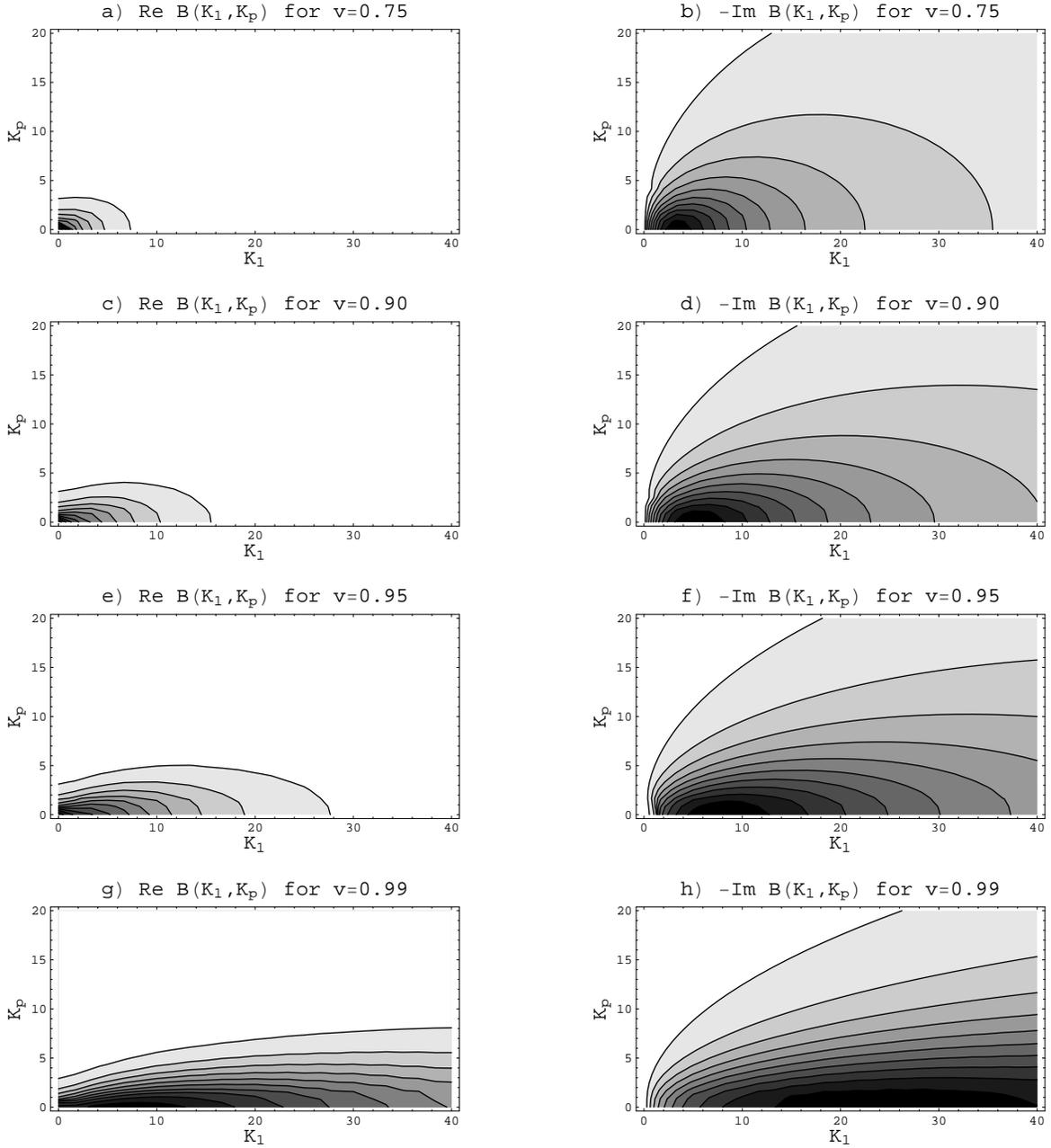}}
  \caption{Contour plots of the real part and minus the imaginary part of $B(K_1,K_\perp)$ for several values of $v$.  The near field contribution \eno{FoundBKagain} has been subtracted.  $B(K_1,K_\perp)$ is proportional to the $K$-th Fourier mode of $\langle {\cal O}_{F^2} \rangle$: see \eno{GotO}.  In each plot, the white region is closest to zero, and the black region is the most positive.}\label{Unsubtracted}
 \end{figure}

We take advantage of the axial symmetry of the problem to express
$B_K = B(K_1,K_\perp)$ where $K_\perp = \sqrt{K_2^2 + K_3^2}$.
Because $\phi(t,\vec{x},z)$ and $\langle {\cal O}_{F^2}(t,\vec{x})
\rangle$ are real, it must be that $B(-K_1,K_\perp) = B(K_1,K_\perp)^*$.  It is easy to see that
this condition is enforced by the differential equation.  Our
results for $B(K_1,K_\perp)$, with the near field \eno{FoundBKagain} subtracted, are shown in figures~\ref{Unsubtracted} and~\ref{AbsValues}.  A good match to the near field form \eno{FoundBKagain} was obtained: for $K_\perp > 10$ the deviations are at the level of tenths of a percent.  These deviations are interesting and can be seen in magnified form in panes b, d, f, and~h of figure~\ref{AbsValues}.  Much of our discussion in section~\ref{DISCUSSION} will hinge on these high-momentum tails.

\section{Discussion}
\label{DISCUSSION}

Before attempting a comparison of our results with recent RHIC results, we will give a brief summary of how the measurements of interest are done.  The reader is warned that we are non-experts and is referred to the experimental literature---for example \cite{Adler:2002tq,Wang:2004kf,Adler:2005ee}---for an authorative account.

Consider the following scenario:
 \begin{enumerate}
  \item Two highly energetic partons collide near the surface of the hot dense matter produced in a relativistic heavy ion collision.  After the collision, the partons have large transverse momentum.
  \item One parton escapes without interacting significantly with the quark-gluon plasma (QGP) and fragments in vacuum into what is termed the near side jet.
  \item The other parton travels through the QGP.  Its evolution into observed particles is strongly affected by its interaction with the QGP.  If it weren't for these interactions, this parton would simply fragment into an away side jet, approximately back-to-back with the near side jet.
 \end{enumerate}
Because of difficulties in unambiguously identifying jets, a standard strategy is to look for angular correlations between two energetic charged particles: the trigger particle, which is presumed to be part of the near-side jet, and the partner particle, which is the putative probe of jet-quenching.  Histograms of the azimuthal angle $\Delta\phi$ between these two particles invariably show a peak at small angles, which means that the partner particle is often part of the near-side jet.  A peak at $\Delta\phi=\pi$ is evidence for an away side jet.  In central collisions, the peak at $\Delta\phi=\pi$ disappears \cite{Adler:2002tq} or even splits \cite{Wang:2004kf,Adler:2005ee}.  In \cite{Adler:2005ee}, the trigger particle is required to have $2.5\,{\rm GeV}/c < p_T < 4.0\,{\rm GeV}/c$ while the partner particles has $1.0\,{\rm GeV}/c < p_T < 2.5\,{\rm GeV}/c$, and for central collisions a broad peak is observed roughly between $\Delta\phi=1.6$ and $\Delta\phi=2.6$.  (All angles will be quoted in radians.)  There is actually a minimum at $\Delta\phi = \pi$.

The recent theoretical literature on jet-quenching, with which we
have less familiarity than we would like, offers several
possibilities.  Among them are scenarios
\cite{Greco:2003xt,Hwa:2004ng} where the QGP affects fragmentation
by recombination of thermal quarks with the parton shower;
extensions of traditional QCD methods such as the twist expansion
\cite{Majumder:2004pt}; predictions of a coherent high momentum
ridge of color flux emanating from the quark
\cite{Fries:2004hd,Armesto:2004pt}; and related discussions of a QCD
``sonic boom'' giving rise to conical collective flow
\cite{Casalderrey-Solana:2004qm,Ruppert:2005uz}.

In the backdrop of these experimental and theoretical investigations, it is interesting to say what we can about the energy flow and spectrum of particles radiated from the heavy quark described in the previous sections.  The hazards of comparing strongly coupled ${\cal N}=4$ super-Yang-Mills with real-world QCD are well known: for a brief summary, see \cite{Gubser:2006bz}.  To these difficulties we must add that we have treated the quark as infinitely massive, whereas the experimental results we have referred to do not include heavy-quark tagging.  Also, it would be better to know $\langle T_{\mu\nu} \rangle$ in addition to $\langle {\cal O}_{F^2} \rangle$: energy flow is most crisply captured in the Poynting vector $S_i = T_{0i}$.  Finally, it would be desirable to go to larger $K_1$ and $K_\perp$, which requires either CPU-intensive numerics or an improved calculational method, as discussed near the end of section~\ref{NUMERICS}.

Objects deep inside $AdS_5$ are understood to correspond to soft
field configurations in the dual CFT, while objects near the
boundary correspond to more localized configurations.  So a
reasonable expectation based on \cite{Herzog:2006gh,Gubser:2006bz}
is that the profile of $\langle {\cal O}_{F^2} \rangle$ would have
the form of a wake, consistent with the ideas of
\cite{Fries:2004hd,Armesto:2004pt,Casalderrey-Solana:2004qm,Ruppert:2005uz}.
The Fourier space profiles shown in figures~\ref{Unsubtracted}
and~\ref{AbsValues} suggest a slightly different, possibly complementary picture.  It helps our intuition to use explicit
numbers.  Let's set
 \eqn{SetT}{
  T = {1 \over \pi}\,{\rm GeV} = 318\,{\rm MeV} \,.
 }
This is in the upper range of temperatures for the QGP, and it is a convenient choice for us because the $K_1$ and $K_\perp$ axes in figure~\ref{Unsubtracted} and~\ref{AbsValues} can then be read in units of ${\rm GeV}/c$.  Another interesting number is the typical final energy of a free massless particle that collides elastically with the heavy quark.  To compute this we take the initial momentum of the massless particle to be of magnitude $T$ and directed perpendicular to the heavy quark's velocity.  If the perpendicular component of the massless particle's momentum doesn't change during the collision, then its final energy is
 \eqn{TGRE}{
  E_f = {1+v^2 \over 1-v^2} T
      = 6.2\,{\rm GeV}  \qquad\hbox{for $v=0.95$.}
 }
We have indicated $E_f$ for the various velocities with the green dots in panes a, c, e, and~g of figure~\ref{AbsValues}.  If the gauge theory were almost free instead of strongly coupled, we would expect the energy loss to be dominated by collisions of the type that led to \eno{TGRE}.

For $v=0.95$, $|B(K_1,K_\perp)|$ is peaked in a range of momenta
between $2$ and $7\,{\rm GeV}/c$ (the black region in
figure~\ref{AbsValues}).  Because ${\cal O}_{F^2} \sim \tr F^2$
starts with bilinears in the fundamental fields, this would
correspond to radiated particles with momenta between $1$ and
$3.5\,{\rm GeV}/c$: less than the $E_f$ of \eno{TGRE} by a factor of
a few.  For $v=0.99$, half the momentum at which $|B(K_1,K_\perp)|$
is peaked is less than $E_f$ by a similar factor.  These
considerations encourage the view that dissipative events involve
several quanta interacting with each other as they recoil from the
heavy quark.  This is broadly consistent with the picture of a
coherent co-moving high momentum ridge dissipating energy from the
heavy quark. But as we will see below, the peak regions of
$|B(K_1,K_\perp)|$ may not dominate the dissipative physics.

Panes b, d, f, and~h of figure~\ref{AbsValues} show that if one
multiplies $|B(K_1,K_\perp)|$ by the factor $K_\perp$ that would
arise in an integration over momentum space, the result is
directionally peaked.  This again brings to mind the picture of
dissipation through radiation carried mostly in the high momentum
ridge. The opening angle $\theta$ between the heavy quark's velocity
and the directional peak of $K_\perp |B(K_1,K_\perp)|$ depends
strongly on the speed:
 \eqn{ThetaTable}{
  \begin{tabular}{c|cccc}
   $v$ & $0.75$ & $0.90$ & $0.95$ & $0.99$ \cr\hline
   $\theta$ & $0.58$ & $0.41$ & $0.30$ & $0.17$
  \end{tabular}
 }
The values of $\theta$ in \eno{ThetaTable} were determined by
setting the $K_\perp$ derivative of $K_\perp |B(K_1,K_\perp)|$ to zero at
fixed and large $K_1$, then taking the appropriate arctangent
function to find $\theta$.

The phase space factor $K_\perp$ makes an enormous difference to the dominant momentum scale.  In $K_\perp |B(K_1,K_\perp)|$, momenta many times $E_f$ dominate.  Indeed, along the preferred direction, $K_\perp |B(K_1,K_\perp)|$ seems to level off at a finite value as $K$ increases.  Evidently we have not explored sufficiently high momenta to discern whether the region where $K_\perp |B(K_1,K_\perp)|$ is above a finite threshhold has finite volume.\footnote{Note that in the large momentum region of the plots shown, we are subtracting a quantity, $B^{\rm near\ field}_K$, which scales linearly with momenta.  The remainder, $B(K_1,K_\perp)$, scales roughly as $1/K$ in the region in question.  This evidently requires substantial numerical precision.  All internal checks of our numerical results suggest they are robust, but the importance of large $K$ tails to our discussion is the reason we say it would be value to have WKB methods in hand.}

To recap: the plots of $K_\perp |B(K_1,K_\perp)|$ not only indicate
directionality, but also suggest that highly energetic fields are an
important part of the description of the radiation process.  To
appreciate just how energetic, note that a charm quark moving in
vacuum with $v=0.95$ has energy $4.5\,{\rm GeV}$, while a $b$ quark
with this speed has energy $15\,{\rm GeV}$.  If $K_\perp
|B(K_1,K_\perp)|$ can be used as an approximate guide to the
spectrum of radiated particles, the single particle energy could
easily be in the $10\,{\rm GeV}$ ballpark.  Recoil would obviously
become an important consideration if a real-world $c$ or $b$ quark
emitted a particle even approaching this range.  This would
substantially increase the opening angle $\theta$.  And it would
encourage the idea that the QGP enhances fragmentation processes at
energies close to the kinematic limit.

There are two main reasons to treat with particular caution a ``prediction'' from AdS/CFT that heavy quarks should undergo fragmentation near the kinematic limit:
 \begin{enumerate}
  \item We have not made a quantitatively precise connection between $\langle {\cal O}_{F^2} \rangle$ in Fourier space and the spectrum of radiated particles.  Indeed, the peak region of $B(K_1,K_\perp)$ and its high-momentum tails send conflicting messages about the spectrum.  We believe the tails are important, but it may be that they have to do mostly with fields near the quark rather than radiative dynamics.  The question of the spectrum of radiated particles should be revisited purely within the context of AdS/CFT with the VEV of the stress-energy tensor in hand, and preferably with semi-analytic methods to buttress numerical analysis of the high-momentum tails.\footnote{Indeed, a computation of the stress tensor gives clear-cut evidence at smaller wave-numbers for a wake in sense usually meant by phenomenologists \cite{Friess:2006fk}.}
  \item Relating hard processes in strongly coupled ${\cal N}=4$ super-Yang-Mills and QCD is especially perilous.  Elementary scattering processes with large momentum transfer can be treated perturbatively in QCD.  In strongly coupled ${\cal N}=4$ super-Yang-Mills the general expectation is that they cannot.  But one should bear in mind that many amplitudes of ${\cal N}=4$ are protected against all loop corrections.  It would be interesting to inquire whether amplitudes for gluons scattering off an external quark have non-renormalization properties.  This discussion recalls the basic conundrum of the connection between AdS/CFT and RHIC: are near-extremal D3-branes merely an analogous system to the QGP, or can they capture the dynamics of real-world QCD above the confinement transition sufficiently precisely to be a useful guide to RHIC physics?
 \end{enumerate}

Fragmentation near the kinematic limit seems to us consistent with the broad peak in $\Delta\phi$ observed in \cite{Adler:2005ee}.  But the energy ranges for the hadrons in \cite{Adler:2005ee} are substantially lower, relative to the temperature, than the energies we have discussed in relation to AdS/CFT.  Recall that the upper limit on $p_T$ of the partner particle is $2.5\,{\rm GeV}/c$.  If the typical energy of the partner particles is sufficiently low, it would be a blow to the picture of enhanced high-energy fragmentation.  Of course, without tagging most of the partons studied in \cite{Adler:2005ee} may be presumed to be light quarks or gluons.

In summary, the calculations we perform are based on the trailing
string picture of \cite{Herzog:2006gh,Gubser:2006bz}, which naively
supports the notion of a coherent wake of color fields with the
heavy quark at its tip.  We do find evidence for a directional ``prow,'' which becomes more and more forward as the speed increases.  It seems that a full description of this prow involves high-momentum
gauge fields.  This may be a hint that, with a realistic cutoff on
the quark mass imposed by hand, the quark could be deflected
significantly by a single radiative event.

The drag force \eno{DragForce} computed in \cite{Herzog:2006gh,Gubser:2006bz} is a time-averaged quantity which provides no direct information about the energy scale of radiated particles.  Calculating color-singlet VEV's in the boundary theory gives considerably more detailed information.  Despite the hurdles string theory faces in connecting to relativistic heavy ion collisions, we hope that the trailing string picture can be further exploited to understand energy loss in the QGP.

\section*{Acknowledgments}

We thank W.~Zajc for comments on an early draft.  This work was supported in part by the Department of Energy under
Grant No.\ DE-FG02-91ER40671, and by the Sloan Foundation.  The work
of J.F.~was also supported in part by the NSF Graduate Research
Fellowship Program.

\bibliographystyle{ssg}
\bibliography{wake}

\end{document}